\documentclass[twocolumn,twoside,slac_two]{revtex4}
\usepackage{graphicx}
\usepackage{fancyhdr}
\begin{document}

\title{A Unified Model for Gamma-Ray Bursts}

%

\author{Ryo Yamazaki}
\affiliation{Department of Earth and Space Science,
Osaka University, Toyonaka 560-0043, Japan}
\author{Kunihito Ioka}
\affiliation{Physics Department and Center for Gravitational Wave Physics,
104 Davey Laboratory, Pennsylvania State University, University Park,
PA 16802, USA}
\author{Takashi~Nakamura, Kenji~Toma}
\affiliation{Theoretical Astrophysics Group
Department of Physics, Kyoto University
Kyoto 606-8502, Japan}

\begin{abstract}
A possible unified model of short and long gamma-ray bursts
(GRBs), X-ray rich GRBs, and X-ray flashes is proposed.
It is assumed that the jet of a GRB  consists of many emitting
sub-shells (i.e., an inhomogeneous jet model).
The multiplicity of the sub-shells along a line of sight $n_s$ is 
an important parameter. 
If $n_s$ is large ($\gg 1$) the event looks like a long GRB, 
while if $n_s=1$, the event looks like a short GRB. 
Finally, when $n_s=0$, 
the event looks like an X-ray flash or an X-ray rich GRB.
Furthermore, our model may also explain the
 bimodal distributions of $T_{90}$ duration of BATSE-GRBs.
Clearly,  our model predicts that short GRBs should be associated
with energetic SNe.
\end{abstract}

\maketitle

\thispagestyle{fancy}


\newcommand{\tot}{{\rm tot}}
\def\N{\nonumber}
\def\f{\frac}
\def\ga{\gamma}
\def\max{{\rm max}}
\def\min{{\rm min}}
\def\j{{\scriptscriptstyle (j)}}
\def\tot{{{\rm tot}}}
\def\sub{{{\rm sub}}}
\def\obs{{{\rm obs}}}
\def\dep{{{\rm dep}}}

\section{Introduction}

For the long gamma-ray bursts (GRBs), 
the cosmological distance, the collimated jet,
the massive star progenitor, and the association with the supernova 
are almost established or strongly suggested
\cite{piran99,m02,zm03}.
However,  for  short GRBs, little is known
since no afterglow has been observed.
Observed bimodal distribution of $T_{90}$ duration implies
the origin of the short burst is different from that of long GRB,
e.g., coalescencing  binary neutron stars.
The origin of the X-ray flashes (XRFs) also remains
unclear although many models have been proposed
(see \cite{yin04a} and references therein).
The observed event rate of  short GRBs and XRFs are
 about a third of and similar to the long GRBs, respectively 
\cite{He01a,ki02,lamb2003}.
Although there may be a possible bias effect to these statistics, 
in an astrophysical sense, these numbers are the same or  comparable. 
If these three phenomena arise from essentially different origins, 
the similar number of events is just by chance. 
While if these three phenomena are related like 
a unified scenario for active galactic nuclei
\cite{awaki1991,antonucci1993,urry1995},
the similar number of  events is natural and 
the ratio of the event rate tells us something about
the geometry of the central engine.
We propose a unified model in which
the central engine of short GRBs, long GRBs and  XRFs is the same  
and the apparent differences come essentially from  different viewing angles. 
This paper is a slight extention of ref~\cite{yin04b}
(see also ref~\cite{toma2004}).

\section{Unified Model}

It is suggested that  short GRBs are similar to 
the first 1 sec of  long GRBs \cite{ggc03}.
Although  short GRBs are harder than long GRBs \cite{k93},
this difference is mainly due to the difference 
in the low-energy spectral slope while the peak energy is similar
\cite{ggc03}.
Fine temporal structures in observed light curves of short GRBs
are similar to those of the first several seconds of 
long GRBs \cite{nakar2002}.
Other properties, such as $\langle V/V_{\rm max}\rangle$, 
the angular distribution,
the energy dependence of duration
and the hard-to-soft spectral evolution
of  short GRBs, are also similar to those of long GRBs
\cite{lamb2002}.
If  short GRBs also obey the peak energy-luminosity relation 
found for the long GRBs \cite{y03},
it is suggested that  short and long GRBs have
a similar redshift distribution \cite{ggc03}.

\begin{figure}[t]
\centering
\includegraphics[width=8cm]{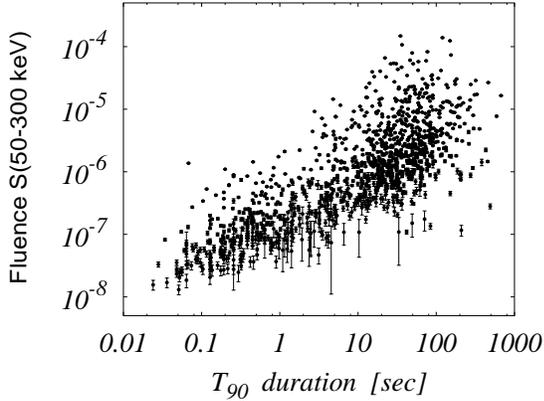}
\caption{
The fluence $S(50-300~{\rm keV})$ as a function of $T_{90}$
duration for BATSE bursts from 4Br catalog
(Courtesy of Drs.~S.~Michikoshi and T.~Suyama). \ From
ref~\cite{yin04b}
}
\label{fig_flu_dur}
\end{figure}

These similarities suggest that the difference between  short 
and long GRBs is just the number of pulses, 
and each pulse is essentially the same \cite{rf00}.
As shown in Fig.~\ref{fig_flu_dur},
using 4Br catalogue of BATSE \cite{pac99}, the fluence is roughly 
in proportion to the duration in the range of 0.01 to 1000 sec
\cite{balazs2003}.
Thus, we may consider that each pulse is produced
by essentially the same unit or the sub-jet\footnote{
The origin of sub-jets (or equally emitting sub-shells) is discussed in
\S~4.1.}, and the GRB jet consists 
of many sub-jets.
If many sub-jets point to our line of sight, the event looks like 
the long GRB while  if a single sub-jet points to us, 
the event looks like a short GRB.
Since we can observe only the angular size of $\sim \gamma^{-1}$
within the GRB jet with the Lorentz factor $\gamma$,
different observers will see different number of sub-jets
depending on the distribution of sub-jets within the GRB jet.
Since the angular size of a causally connected region is also
$\gamma^{-1}< 0.01 $,
the opening half-angle of a sub-jet can be much smaller than that of
the whole GRB jet ($\sim 0.1$), say $\sim0.02$.

XRFs also appear to be related to GRBs.
Softer and dimmer GRBs smoothly extend to the XRFs
\cite{He01a,ki02,lamb2003,watson04},
while the peak energy-isotropic luminosity/energy relations hold
for GRBs as well as XRFs \cite{s03,y03,a02}.
%
The total energy including the radio afterglow of 
XRF~020903, which has a measured redshift,
might be similar to that of  GRBs \cite{so03}.
Other properties, such as the duration, the temporal structure
and the Band spectrum of the XRFs are also similar to those of the GRBs,
suggesting that XRFs are in fact soft and dim GRBs.
In the sub-jet model, XRFs are naturally expected 
when our line of sight is off-axis to any sub-jets
\cite{nakamura2000,in01,yin02,yin03b,yin04a}.

\section{An Example of Numerical  Simulation of Our Unified Model}

\begin{figure}[t]
\centering
\includegraphics[width=8cm]{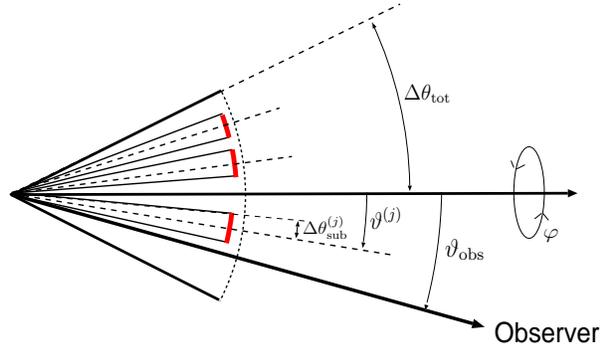}
\caption{Schematic diagram of the multiple subjet model.
Each subjet with opening half-angle $\Delta\theta_\sub$ is launched
within a cone with an opening half-angle 
$\Delta\theta_\tot$\,. From ref~\cite{yphd}.
}
\label{fig:subjets}
\end{figure}

\begin{figure}[t]
\centering
\includegraphics[width=8cm]{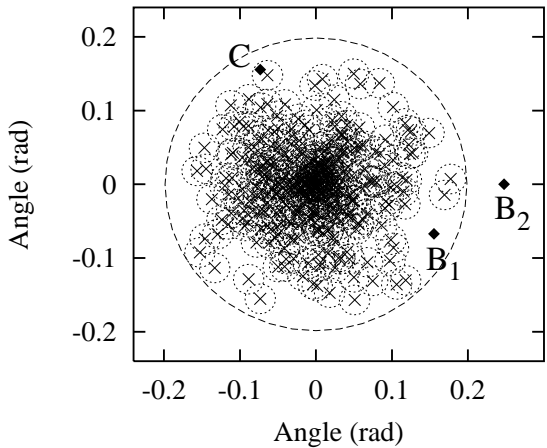}
\caption{
The angular distribution of $N_\tot=350$ sub-jets confined 
in the whole GRB jet in our simulation.
The whole jet has the opening half-angle of 
$\Delta\theta_\tot=0.2$~rad.
The sub-jets have the same intrinsic luminosity,
opening half-angles $\Delta\theta_\sub=0.02$~rad and 
other properties; $\gamma=100$, $r=10^{14}$\,cm, 
$\alpha_B=-1$\,, $\beta_B=-2.5$\,, $h\gamma\nu'=500$\,keV.
The axes and the angular size of sub-jets are represented by crosses
and the dotted circles, respectively. 
 ``A'' represents the center of the whole jet and is hidden by the lines
 of sub-jets. \ From
ref~\cite{yin04b}.
}\label{fig1}
\end{figure}

In the following, we show a numerical simulation to demonstrate how
an event looks so different depending on the viewing angle 
in our unified model \cite{yin04b,yphd}.
Let us consider $N_\tot=350$ sub-jets, for simplicity, 
confined in the whole GRB jet
whose axis is the same as a $\vartheta=0$ axis (see fig.~\ref{fig:subjets}).
For each sub-jet the emission model is the same as in \cite{yin03b}.
Let the opening half-angle of the $j$-th sub-jet 
($j=1,\,\cdots,\,N_\tot$) be $\Delta\theta_\sub^\j$,
while the opening half-angle of the whole jet be
$\Delta\theta_\tot$.
The direction of the observer and the axis of the $j$-th sub-jet 
are specified by $(\vartheta_\obs\,,\varphi_\obs)$
and $(\vartheta^\j,\varphi^\j)$, respectively.
We assume the $j$-th sub-jet departs at time $t_\dep^\j$
from the central engine
and emits at radius $r=r^\j$ and time 
$t=t^\j\equiv t_\dep^\j+r^\j/\beta^\j c$,
where $t$ and $r$ are measured in the central engine frame
and we set $t_\dep^{\scriptscriptstyle (j=1)}=0$.
For simplicity, all sub-jets are assumed to have the same intrinsic
properties, that is 
$\Delta\theta_\sub^\j=0.02$~rad, $\gamma^\j=100$~, 
$r^\j=10^{14}$~cm,
$\alpha_B^\j=-1$, $\beta_B^\j=-2.5$, 
$\gamma h{\nu'}_0^\j=500$~keV and the amplitude 
$A^\j={\rm const}.$ for all $j$.
The departure time of each sub-jet, $t_\dep^\j$ is randomly
distributed between $t=0$ and $t=t_{\rm dur}$,
where $t_{\rm dur}$ is the  active time
of the central engine measured in its own frame and set to
$t_{\rm dur}=30$~sec.
The opening half-angle of the whole jet is set to 
$\Delta\theta_\tot=0.2$~rad as a typical value.
We consider the case in which 
the angular distribution of  sub-jets is given by 
\[
P(\vartheta^\j,\varphi^\j)\,d\vartheta^\j\, d\varphi^\j\propto
\exp[-(\vartheta^\j/\vartheta_c)^2/2]
\,d\vartheta^\j\, d\varphi^\j
\]
for $\vartheta^\j<\Delta\theta_\tot-\Delta\theta_\sub$,
where we adopt  $\vartheta_c=0.1$~rad \citep{z03}.
In this case,
sub-jets are concentrated on the $\vartheta=0$ axis (i.e., the
multiplicity in the center $n_s\sim 10$).
For our adopted parameters,  sub-jets are sparsely distributed in the 
range $\vartheta_c\lesssim\vartheta\lesssim\Delta\theta_\tot$,
however,  the whole jet would be entirely filled
if the sub-jets were uniformly distributed 
(i.e., the mean multiplicity $n_s\sim 3$).
Therefore, isolated sub-jets exist near the edge of the whole jet 
with the multiplicity $n_s\ll 1$
and there exists a viewing angle where no sub-jets are launched.
Figures~\ref{fig1}, \ref{fig2} and \ref{fig3} show 
the angular distributions of sub-jets and
the directions of four selected lines of sight, the observed time-integrated
spectra, and the observed light curves in the X-ray and $\gamma$-ray
bands, respectively. Note here in Figure~\ref{fig1}, ``A'' represents 
the center of the whole jet and is hidden by the lines of sub-jets.


\subsection{Long GRB}
When we observe the source from the $\vartheta=0$ axis
(case~``A''), we see spiky temporal structures
(the upper-middle panel of Fig.~1)  and  $E_p \sim 300$~keV
which are typical for the long GRBs.
We may identify  case ``A'' as  long GRBs. 

\subsection{XRF and X-ray rich GRB}

When the line of sight is away from any sub-jets
(cases~``B$_1$'' and ``B$_2$''),
soft and dim prompt emission, i.e.  XRFs
or X-ray rich GRBs are observed with $E_p= 10\sim 20$~keV and $\sim 4$
orders of magnitude smaller fluence than that of   case ``A''
(Fig. 2). 
The burst duration is comparable to that in  case  ``A''. 
These are quite similar to the characteristics of XRFs 
\cite{nakamura2000,in01,yin02,yin03b,yin04a}.
We may identify the
cases ~``B$_1$'' and ``B$_2$'' as  XRFs or X-ray rich GRBs.

In our previous works \cite{yin02,yin03b,yin04a},
we considered  the homogeneous, instantaneous emission of the
whole jet. Then XRFs and X-ray rich GRBs occur only when the whole jet is
viewed off-axis (corresponding case~B$_2$).
We now introduce sub-structure of the jet emission. Then,
there exists the case when the observer sees all sub-jets off-axis but
his line of sight is within the whole jet (case~B$_1$).
This leads two types of X-ray afterglows because in the afterglow phase,
the viewing angle of the whole jet becomes important.
For details see section~4.3.

\subsection{Short GRB}
If the line of sight is  inside an isolated sub-jet (case~``C''),
its observed pulse duration is $\sim 50$ times smaller than 
 case ``A''.
Contributions to the observed light curve from the other sub-jets
are negligible, so that the fluence is  about a hundredth
 of the case ``A''. 
These are quite similar to the characteristics of  short GRBs.
However the hardness ratio 
($=S(100-300~{\rm keV})/S(50-100~{\rm keV})$)  is about 3 
which is smaller than the mean hardness of short GRBs ($\sim 6$). 
It is suggested that the hardness of  short GRBs
is due to the large low-energy photon index $\alpha_B\sim -0.58$ 
\cite{ggc03} 
so that if the central engine
launches $\alpha_B\sim -0.58$ sub-jets to the periphery of the core 
where
 $n_s$ is small, we may identify the case ``C'' as the short-hard GRBs.
In other words, the hardness of 3 comes from $\alpha_B =-1$ in our simulation
so that if $\alpha_B\sim -0.58$, the hardness will be 6 or so. 
We suggest here that not only the isotropic energy but also the photon
index may depend on $\vartheta$. Another possibility is that
if short GRBs are the first 1~sec of the activity of the 
central engine, the spectrum in the early time might be
$\alpha_B\sim -0.58$ for both  the sub-jets in the core and the 
envelope. 
This is consistent with a high KS test probability for $E_p$  and 
$\alpha_B$ \cite{ggc03}. These possibilities may
have something to do with the origin of $\alpha_B\sim -1$ 
for the long GRBs. 

\subsection{X-ray pre-cursor/post-cursor}
It is quite interesting that in Figure~\ref{fig3},
we see the X-ray precursor at $T_{\rm obs}\sim60$~sec in  ``B$_2$'' 
 and the postcursor at $T_{\rm obs}\sim65$--75~sec in ``B$_1$''.
These can be understood by the model proposed by \cite{nakamura2000}.

\begin{figure}[t]
\centering
\includegraphics[width=7cm]{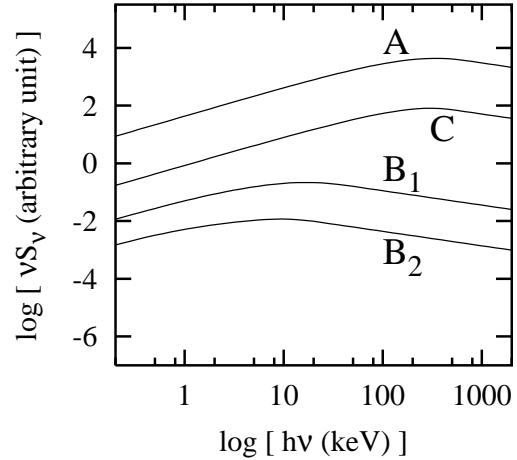}
\caption{
Time-integrated energy spectrum of the emission from the multiple sub-jets
for the observers denoted by 
``A'', ``B$_1$'', ``B$_2$'', and ``C'' in Figure~\ref{fig1}.
The source are located at $z=1$. \ From
ref~\cite{yin04b}.
}\label{fig2}
\end{figure}

\begin{figure*}[t]
\centering
\includegraphics[width=15cm]{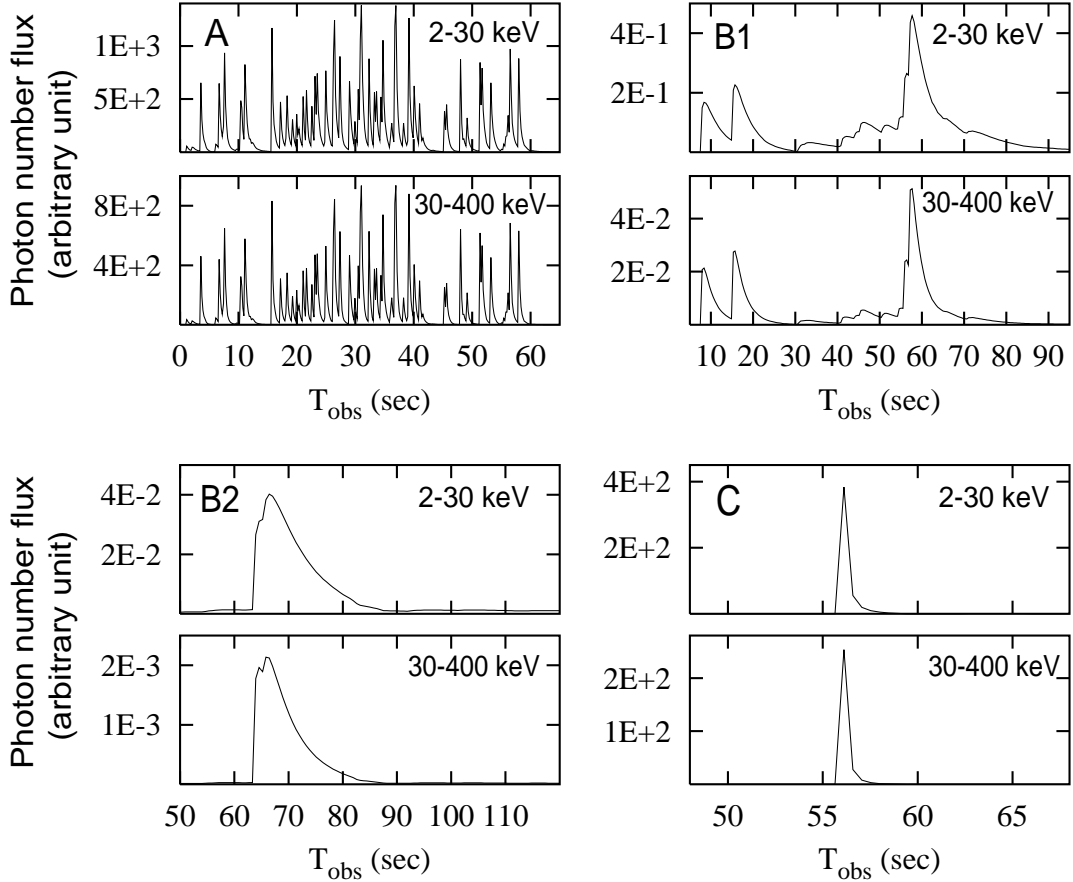}
\caption{
The observed X-ray and $\gamma$-ray light curves from the multiple 
sub-jets,
corresponding the cases ``A''(the upper left),
``B$_1$''(the upper right), ``B$_2$''(the lower left) and
``C''(the lower right) in Figure~\ref{fig1}.
The sources are located at $z=1$. \ From
ref~\cite{yin04b}.
}\label{fig3}
\end{figure*}

\section{Discussions}

\subsection{Jet structures}

\underline{\it Origin of Sub-jets.} --- 
In this paper, we do not discuss the origin of the sub-jets,
but argue the implications of the multiple sub-jet model
\cite{kp00,nakamura2000}
This  model is an anologue of the
 patchy shell model for an afterglow \cite{np03,png03,no04}.  
Note that in our model, a relativistically moving emitting sub-shell
is called a sub-jet.
The origin of sub-jets is not yet clear.
One possibility is that they may arise from relativistically 
outflowing blobs generated by various fluid instabilities like
Kelvin-Helmholtz, Rayleigh-Taylor instability, and so on
\cite{aloy2002,zwh03,gomez2003}.

Jet structures (sub-jet configurationa) could be determined 
in order to reproduce all of the
observed statistical properties even including those of afterglows
or gravitational waves \cite{siny04}.

\underline{\it The Number of Isolated Sub-jets.} --- 
Let $\Delta\theta_{\rm sub}$, $\vartheta_c$ and $\bar{n}_s$ be the
typical opening half-angle of the sub-jet, the core size of the whole jet
and the mean multiplicity in the core. Then the total number of the
sub-jets ($N_\tot$) is estimated as
$N_\tot=\bar{n}_s(\vartheta_c/\Delta\theta_{\rm sub})^2\sim 10^3$,
so that the total energy of each sub-jet is $\sim 10^{48}$ erg. 
In our model, 
the event rate of long GRBs is in proportion to $\vartheta_c^2$. 
Let $M$ be the number of sub-jets in the 
envelope of the core with a  multiplicity $n_s=1$. 
Then the event rate of short GRBs is in proportion to
$M\Delta\theta_{\rm sub}^2$, so that $M\sim 10$ is enough to explain 
the event rate of short GRBs. 

\underline{\it Angular Distribution of Sub-jets.} --- 
Of course, the above numerical values are typical ones
and should have a dispersion \cite{ldz03}.
Our core-envelope sub-jet model can have a similar structure to
the two component jet model 
\cite{b03,h03,zwh03,rcr02}
by varying such as  $\bar{n}_s$ and $M$.
However the distribution of sub-jets could also have other 
possibilities, e.g., a hollow-cone distribution like a pulsar, 
a power law distribution, a Gaussian distribution 
\cite{zm02,rossi2002,z03} and so on.

\subsection{Properties of Prompt Emission}

\begin{figure}[t]
\centering
\includegraphics[width=8cm]{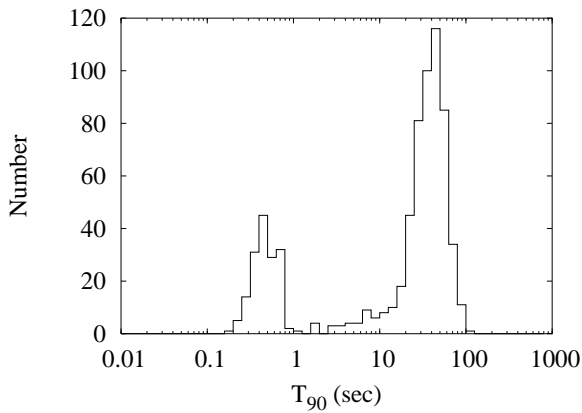}
\caption{
$T_{90}$ duration distribution in 50--300~keV of hard events
with observed fluence ratio 
$S(2-30~{\rm keV})/S(30-400~{\rm keV}) < 10^{-0.5}$.
Subjet distribution is given by Gaussian form.
The source redshifts are varied according to the cosmic star formation
rate.
}
\label{fig5}
\end{figure}

\underline{\it Bimodal Distribution of $T_{90}$ Duration.} --- 
It is also found that our model can reproduce the bimodal
distribution of $T_{90}$ duration of GRBs observed by BATSE \cite{toma2004}.
In our model,
the duration of $n_s=1$ burst is determined by the angular
spreading time of one sub-jet emission, while
that of $n_s \geq 2$ burst is determined by the time interval between
the observed first pulse and the last one
(see also Fig.~2 of ref~\cite{toma2004}).
These two different time scales naturally lead a division of the burst
$T_{90}$ durations
into the short and long ones.
We show an example in Figure.~\ref{fig5}.
The dispersion of the lognormal-like distribution seems relatively small
compared with the observations.
This is ascribed to a simple modeling in this paper.
We fix the jet configuration and
use the same intrinsic properties of the subjets.
If we vary $t_{\rm dur}$ for each source and
$\gamma^{\j}$ for each subjet randomly, for example,
the dispersion of lognormal-like $T_{90}$ duration distribution
will increase from the general argument that the dispersion of the
lognormal
distribution increases with the increase of the number of the associated
random variables \cite{in02}.
In more realistic modeling
the observed dispersion will be reproduced.

It has commonly been said that
the observed bimodal distribution of $T_{90}$ durations of BATSE bursts
shows the different origins of short and long GRBs.
However, the bimodal distribution is also available as a natural
consequence of
our unified model of short and long GRBs.

\underline{\it Temporal Structures of Long GRBs and XRFs.} --- 
There are three important time scales.
The first is the duration of the central engine
measured by the observer 
$T_{\rm dur}=(1+z)\,t_{\rm dur}$.
The second is the observed pulse duration of $j$-th sub-jet, $\delta T^\j$,
which is given by the angular spreading time scale of each subjet.
Since emission components far from the viewing angle of $\gamma^{-1}$
is dim due to the relativistic beaming effect,
it can be approximated as
\begin{eqnarray}
 \delta T^\j&\sim&(1+z)\f{r^\j}{c}
\Bigl[\cos\tilde{\theta}-\cos(\tilde{\theta}+\gamma^{-1})\Bigr]\nonumber\\
&\sim&(1+z)\f{r^\j}{2c\gamma^2}(1+2\gamma\tilde{\theta})~~,\nonumber
\end{eqnarray}
where $\tilde{\theta}\equiv\max\{0,\theta_v^\j-\Delta\theta_{\rm sub}^\j\}$,
and $\theta_v^\j$ is the viewing angle of the $j$-th sub-jet given by
$\cos\theta_v^\j=\vec{n}_{\rm obs}\cdot\vec{n}_{\rm sub}^\j$ where
unit vectors $\vec{n}_{\rm obs}$ and $\vec{n}_{\rm sub}^\j$ are
specified by directions $(\vartheta_{\rm obs},\varphi_{\rm obs})$
and $(\vartheta^\j,\varphi^\j)$, respectively.
The third is the time $\Delta T$, that is the difference between times 
of arrival at the observer
of photons that arise  simultaneously at the  nearest
and the farthest side of the whole jet to the line of sight, i.e.,
\begin{eqnarray}
\Delta T &\sim&(1+z)\f{r}{c}
\Bigl[\cos(\max\{0,\vartheta_{\rm obs}-\Delta\theta_{\rm tot}\})\nonumber\\
&&   -\cos(\vartheta_{\rm obs}+\Delta\theta_{\rm tot})\Bigr]\nonumber\\
&\sim& (1+z)\f{r}{2c\gamma^2} \nonumber\\
&&
\times\left\{
\begin{array}
{l@{\quad} c@{}}
(\gamma\vartheta_{\rm obs}+\gamma\Delta\theta_{\rm tot})^2, &
\vartheta_{\rm obs}<\Delta\theta_{\rm tot} \\
4(\gamma\vartheta_{\rm obs})(\gamma\Delta\theta_{\rm tot}), &
\vartheta_{\rm obs}>\Delta\theta_{\rm tot} \\
\end{array} \right.~~. \nonumber
\end{eqnarray}
In the following, we assume $z=0$ for simplicity.

Let us consider  the case of $\vartheta_\obs=0$ (then, $n_s\sim10^2$).
Bright pulses in all the sub-jet emissions are observed when
$\theta_v^\j(=\vartheta^\j)\lesssim\gamma^{-1}$. Then
the observed duration of the brightest pulses can be calculated
as $\delta T^\j\sim0.1$\,sec, while
$\Delta T\sim70$\,sec\,, therefore,
$\delta T^\j\ll T_{\rm dur}\lesssim\Delta T$.
Brightest period lasts for about $T_{\rm dur}$.
Since we assume $t_{\rm dep}^\j$ (the departure time of each sub-jet at 
the central engine) is randomly distributed,
 brightest spikes in the observed light curve are
uniformly distributed in this period.
Since the mean time-interval between bright spikes 
 is about $T_{\rm dur}/n_s\sim0.3$~sec and is  larger
than the pulse duration $\delta T^\j$,
those are separated with each other.
This period is followed by that with a duration of  
$\Delta T-T_{\rm dur}\sim40$~sec in which low-flux soft events come from 
sub-jets with large~$\theta_v^\j$.

On the other hand,
when all sub-jets are viewed off-axis, i.e., $n_s=0$,
observed time profile is greatly altered.
In this case,  $\delta T^\j$ increases with the viewing angle and
observed light curves become very smooth.
Since we see the periphery emission,
the number of observed (bright) pulses is small.
For XRFs or X-ray rich GRBs with
$\vartheta_{\rm obs}\sim\Delta\theta_{\rm tot}$
($\vartheta_{\rm obs}\gg\Delta\theta_{\rm tot}$),
the duration of bright emission period is comparable to or 
less than $\sim T_{\rm dur}$ ($\delta T$ for smallest $\theta_v$),
while overall duration is given by $\Delta T$.
There may even be the case when the mean pulse interval in the
observed light curve is smaller than the pulse duration and observed
sub-jet emissions overlap with each other and merge like one or
a few pulses.
Therefore, variability of XRFs and X-ray rich GRBs is low.
This is a possible explanation of observed 
$E_p$--variability relation \cite{lr02}.
More quantitative analysis is under investigation \cite{ytin05}.

\underline{\it Fine Temporal Structures of Short GRBs.} --- 
It has been known that a large fraction of short bursts are
composed of a few pulses with a  duration of several tens of
msec \cite{nakar2002}.  
At present  we use a very simple model of a subjet, i.e,  
instananeous emission at a  certain radius.
In reality there will be a distribution of Lorentz factor of subjet
so that the faster one  will collide to the slower ones.  
Observed variability of short bursts may arise  in such a more 
comlicated modeling of subjet emissions.

\underline{\it Other Properties.} --- 
Some observers could see a cold spot with small $n_s$ in the core
to have a small geometrically corrected
energy even if the total energy of the GRBs is the same.
Thus our model may be compatible with the recent claim
that the total kinetic energy has smaller dispersion
than the geometrically corrected $\gamma$-ray energy 
\cite{b03,bfk03}.
The X-ray pre-/post-cursor is also expected
if off-axis sub-jets are ejected earlier (for precursor)
or later (for postcursor) than the main sub-jets \cite{nakamura2000}.
The viewing angle of the sub-jets may also cause
the luminosity-lag/variability/width relations of the GRBs 
including GRB 980425
\cite{yyn03,in01}.

\subsection{Afterglow Properties}

\underline{\it Early Afterglows.} --- 
At an early phase, the afterglow variabilities
may arise from the angular energy fluctuations within the GRB jet
\cite{np03,png03,no04}, which might correspond to the
inhmogeneous $n_s$.
Indeed, in the context of the patchy shell model,
the observed properties of variable afterglow of GRB~021004 are 
well explained if the angular 
 size of the patches is about 0.02~rad \cite{no04}.
This size is similar to those we adopt in this paper.

\underline{\it Late Phase Afterglows of XRFs.} --- 
The afterglow could have a different behavior between
the core-envelope sub-jet model and the uniform jet model.
In the uniform jet model, the afterglows of XRFs should resemble
the orphan afterglows that initially have a rising light curve
\cite{yin03a,g02}.
An orphan afterglow may be actually observed in XRF 030723 \cite{f04},
but the light curve may peak too early \cite{z03}.
The optical afterglow of XRF 020903 is not observed initially ($<0.9$ days)
but may not be consistent with the orphan afterglow \cite{so03}.
These problems could be overcome by introducing a Gaussian tail
with a high Lorentz factor around the uniform jet \cite{z03}
because the energy redistribution effects may bring the rising light curve
to earlier times \cite{z03,kg03}.
Therefore, as long as the observer points within 
or slightly off-axis to the whole jet
(case B$_1$), the late phase ($\gtrsim1$~day) properties of
XRF afterglow may be similar to those of long GRBs.
On the other hand, 
when the whole jet is viewed far from the edge of the jet,
such that $\vartheta_{\rm obs}\gg\Delta\theta_{\rm tot}$ (case B$_2$),
XRF afterglows may resemble the orphan afterglow
(e.g., see the upper-left panel of fig.~5 in ref~\cite{kg03}.
Afterglows for cases B$_1$ and B$_2$ may correspond $\theta_{obs}=5.7^\circ$
and $\theta_{obs}=11.5^\circ$ in ref~\cite{kg03}, respectively).
Because of the relativistic beaming effect,
case~B$_2$-like events are dimmer than case~B$_1$-like events
 in both the prompt and afterglow phase
(see Figure~\ref{fig2}),
so that they may be rarely observed,
but we believe XRF~030723 is a member of such a class.
Recent calculation for this subject can be found in \cite{grp05}.

\underline{\it Late Phase Afterglows of Short GRBs.} --- 
The afterglow of a short GRB may be difficult to predict since it
could resemble both the orphan and normal afterglow depending on 
the sub-jet configuration within the envelope.
One situation is considered in \cite{fan04},
where global afterglow emission is approximated by that of a
Gaussian structured jet superimposed on a uniform jet.
Then the emission is dominated by the on-axis sub-jet at relatively 
early times, while it is dominated by the more energetic Gaussian core
at later times.
This model assumes the void around the sub-jet that leads
to an afterglow bump.
On the other hand, if the sub-jet structure is produced by one
whole jet through hydrodynamical instability (\S~4.1),
the afterglow of short GRB may be similar to those of structured jet model
because in this case the whole jet is entirely filled with kinetic energy
(i.e., no void around the on-axis sub-jet).

\subsection{Predictions of our Model}

A clear prediction of our unified model is that
short GRBs should be associated with energetic 
SNe\footnote{Indeed, one of short GRBs (GRB~970514) shows the 
possible association with a type~IIn SN (SN~1997cy) \cite{germany00}, 
however, it may be  probable that it was one of
 rare SN~Ia events exploding in a dense circumstellar medium
\cite{turatto2000}.}.
Even if the SNe are not identified with short GRBs due to
some observational reasons we predict that the spatial distribution of
short GRBs in host galaxies should be similar to that of the long GRBs.
Another prediction is that short GRBs should have the same total kinetic
energies as long GRBs, which might be confirmed by radio calorimetry.

Interestingly our model has predicted short XRFs or short X-ray rich
GRBs \cite{yin04b}. 
They are observed when  isolated sub-jets are viewed slightly
off-axis.
The observed short XRF~040924 may be a kind of these bursts
\cite{fenimore04}.
Note that the short XRFs will be longer than the short GRBs
since the pulse duration grows as the viewing angle increases
\citep{in01,yin02}.
The event rate of  short XRFs will depend on the configuration of the
sub-jets in the envelope.
Further observations are necessary to determine the envelope structure.

\subsection{Comment on the NS-NS Merger Model for short GRBs}

Let us assume that short GRBs arises from coalescing binary neutron stars.
Then the current estimate of the coalescing rate is about
$10^{-4}$~y$^{-1}$~galaxy$^{-1}$ \cite{kalogera04} while 
the event rate of long GRBs is estimated as 
$\sim10^{-6}$~y$^{-1}$~galaxy$^{-1}$. If we assume
that the distance to short GRBs is similar to long GRBs, the isotropic energy
of GRBs should be a hundredth of long GRBs because the luminosity of long and 
short GRBs are similar (Figure~\ref{fig_flu_dur})
and the duration of short burst is typically a hundredth of that of
long GRB.           From the event rate, 
the opening angle of the short GRB is an order of magnitude smaller 
than the long GRBs. The typical energy of long GRBs is about 
$10^{51}$~erg and the typical opening half-angle is $\sim0.1$.
This suggests that the total energy of short GRBs is about $10^{47}$~erg,
while $\sim10^{52}$~erg is liberated
from the coalescence of binary neutron star.
This means that the short burst is much less effective and much less 
energetic compared to long GRB.

\bigskip 
\begin{acknowledgments}
R.Y. thanks the conference organizers for inviting him to present this
paper and for partial support.
This work was supported in part by
a Grant-in-Aid for Scientific Research
of the Japanese Ministry of Education, Culture, Sports, Science
and Technology, No.05008 (RY), No.14047212 (TN), and No.14204024 (TN) and
 also  in part by the Eberly Research Funds of Penn State 
and by the Center for Gravitational Wave Physics under grants 
PHY-01-14375 (KI).
\end{acknowledgments}

\bigskip 

\end{document}